\documentclass[12pt]{article}
\usepackage{graphicx}
\usepackage{amsmath,amsthm,amsfonts,amssymb}
\usepackage{color}
\usepackage[normalem]{ulem}
\makeatletter
\@addtoreset{equation}{section}

\makeatother

\begin{document}
	\begin{titlepage}
		\null
		\begin{flushright}
			arXiv:2106.01353
			\\
			June, 2021
		\end{flushright}
		
		\vskip 1.2cm
		\begin{center}
			
			{\LARGE \bf Multi-Soliton Dynamics of \\
				
				\vskip 0.5cm
				
				Anti-Self-Dual Gauge Fields}
			
			\vskip 1.5cm
			\normalsize
			
			{\large 
				Masashi Hamanaka\footnote{E-mail:hamanaka@math.nagoya-u.ac.jp}
				and 
				Shan-Chi Huang\footnote{E-mail:x18003x@math.nagoya-u.ac.jp}
			}
			
			\vskip 0.5cm
			
			{ Department of Mathematics, Nagoya University,\\
				Nagoya, 464-8602, JAPAN}
			
			\vskip 1.5cm
			
			{\bf \large Abstract}
			\vskip 0.5cm
		\end{center}

		We study dynamics of multi-soliton solutions 
		of anti-self-dual Yang-Mills equations for $G=\mathrm{GL(2,\mathbb{C})}$ in four-dimensional spaces. 
		The one-soliton solution can be interpreted as a codimension-one soliton 
		in four-dimensional spaces because the principal peak 
		of action density localizes on a three-dimensional hyperplane. We call it the soliton wall. 
		We prove that in the asymptotic region, 
		the $n$-soliton solution possesses $n$ isolated 
		localized lumps of action density,
		and interpret it as $n$ intersecting soliton walls. 
		More precisely, each action density lump is essentially 
		the same as a soliton wall because it preserves its  
		shape and ``velocity'' except for a position shift of principal peak in the scattering process. 
		The position shift results from the nonlinear interactions of the multi-solitons and is called the phase shift.
		We calculate the phase shift factors explicitly
		and find that the action densities can be real-valued in three kind of signatures. 
		Finally, we show that the gauge group can be $G=\mathrm{SU}(2)$ 
		in the Ultrahyperbolic space $\mathbb{U}$ (the split signature $(+, +, -, -)$). 
		This implies that the intersecting soliton walls 
		could be realized in all region in N=2 string theories. 
		It is remarkable that quasideterminants 
		dramatically simplify the calculations and proofs. 
	\end{titlepage}
	
	\clearpage
	\baselineskip 6.55mm
	
	
	\section{Introduction}
	
	Soliton theories and integrable systems have been studied actively
	and developed rapidly in the past sixty years.
	Meanwhile, the applications of exact solitons, such as instantons, monopoles, vortices, and domain walls also promote the developments in different fields of mathematics and theoretical physics. 
	Therefore, systematic construction of exact soliton solutions has been one of the most attractive topic in the studies of integrable systems.
	Stability of multi-soliton solutions 
	closely relates to the existence of infinite many conserved quantities 
	which leads to an infinite dimensional symmetry of the integrable systems.  
	Among these studies, Sato's theory of solitons is one of the 
	most appealing result which reveals an infinite dimensional symmetry 
	behind the KP equation and gives a comprehensive viewpoint 
	to unify the theory of lower-dimensional integrable systems \cite{Sato}.  
	The key ingredients of Sato's theory are   
	integrable hierarchies and tau functions.
	The tau functions can be represented specifically 
	as Wronskian determinants. This fact is crucial  
	to demonstrate that Hirota bilinear equations are just Pl\"ucker relations 
	by using Maya diagram representations \cite{Sato_lecture,Sato}. 
	Combining them with the integrable hierarchies, 
	we have infinite Pl\"ucker relations which define an 
	infinite dimensional Grassmann manifold 
	as the solution space of the KP equation. 
	In this way, the infinite dimensional symmetry is clarified.
	
	In four-dimensional integrable systems, the most beautiful one 
	would be the anti-self-dual Yang-Mills (ASDYM) equations. 
	The essence of hidden integrability behind the anti-self-dual Yang-Mills 
	equations can be captured clearly under 
	the description of the twistor theory (e.g. \cite{Dunajski,MaWo,PeRi,WaWe}). 
	On the other hand, the anti-self-dual Yang-Mills equations 
	can be reduced to various lower-dimensional soliton equations, 
	such as the KdV equation and the nonlinear Schr\"odinger equation, 
	by suitable reduction procedures \cite{MaWo, Ward}.
	This fact suggests that there might be a profound connection  
	between the Sato's theory and the twistor theory, 
	which perhaps leads to higher-dimensional extension of Sato's theory. 
	A feasible approach to anti-self-dual Yang-Mills equations 
	from the viewpoint of Sato's theory is mentioned in \cite{Takasaki_CMP}, 
	while the description of tau-functions is still unclarified.  
	Another remarkable result is from the viewpoint of 
	B\"acklund transformations \cite{CFGY}. 
	This kind of exact solutions can be represented 
	by determinants in a regular pattern, 
	however, not the Wronskian type determinants. 
	Therefore, the description of tau-functions remains to be clarified.  
	Furthermore, several attempts were made by us to construct 
	one-soliton solutions from \cite{CFGY}, and 
	the resulting action density is Tr$F_{\mu\nu}F^{\mu\nu}=0$ \cite{HaHu}. 
	Perhaps for this reason, only few discussions have been made  
	(as far as the authors know) in this direction for a long time. 
	(e.g. \cite{CSY, deVega, Mason})
	
	Just last year, we had made some progress in this direction. 
	More precisely, we constructed 
	Wronskian type solutions of the anti-self-dual Yang-Mills equations 
	successfully by applying a Darboux transformation \cite{GNO00} 
	in the noncommutative (NC) framework \cite{GHHN}. 
	This Wronskian type solutions can be represented 
	in terms of quasideterminants 
	\cite{GeRe} (called the quasi-Wronskian solutions, for short). 
	A highly nontrivial result of quasi-Wronskian solutions is 
	that the action density in one-soliton case is no longer zero \cite{HaHu}. 
	Moreover, the principal peak of the action density 
	lies on a three-dimensional hyperplane. Therefore, 
	our solutions can be interpreted as codimension-one solitons  
	in four-dimensional space. We call them the soliton walls 
	to distinguish them from the domain walls in this paper.  
	Now a natural question comes: 
	Can we find the behavior of $n$ intersecting soliton walls 
	from the quasi-Wronskian solution? 
	The answer is yes and we verify this by analyzing 
	the asymptotic behavior of $n$-soliton solution \eqref{U2_n} 
	in the scattering process. 
	This guarantees the stability and integrability of 
	the intersecting soliton walls like KP multi-solitons. 
	Moreover, the quasi-Wronskian descriptions are actually 
	more essential than the ordinary Wronskian. 
	This would lead to analogs of the Maya-diagram representation,
	tau-functions, and perhaps Sato's formulations of 
	the anti-self-dual Yang-Mills equations. 
	
	In this paper, we clarify the asymptotic dynamics of multi-soliton solutions 
	of anti-self-dual Yang-Mills equations for 
	$G=\mathrm{GL(2,\mathbb{C})}$  in four-dimensional spaces. 
	We prove that in the asymptotic region, the $n$-soliton solution 
	possesses $n$ isolated localized lumps of action density,
	and interpret it as $n$ intersecting soliton walls. 
	More precisely, each action density lump is essentially the same as 
	a soliton wall because it preserves the shape 
	and ``velocity'' except for a phase shifts. 
	Furthermore, we show that in the Ultrahyperbolic space $\mathbb{U}$, 
	the gauge group can be $U(2)$. This result is important for physical interpretations 
	because the anti-self-dual Yang-Mills equations in this space 
	are equations of motion of effective actions 
	for open N=2 string theories \cite{GOS,Marcus,OoVa}.  
	Therefore, the intersecting soliton walls could be realized 
	in the N=2 string theory in all region of the space-time 
	as new physical objects, that is, intersecting branes. 
	It is remarkable that the quasideterminants play crucial roles 
	in calculations and proofs. 
	As we will comment in the end of section 2 and 
	in the beginning of subsection 4.2, 
	matrix elements of our quasi-Wronskian solutions 
	consist of $2\times 2$ matrices rather than scalar functions. 
	This ``non-abelian treatment'' is quite important 
	to make all proofs and discussions drastically simple
	and expected to be applied to various non-abelian integrable systems.
	
	This paper is organized as follows. 
	In section 2, we make a brief introduction to
	quasideterminants and summarize some properties of them.  
	These are useful mathematical tools for later sections. 
	In section 3, we introduce the $J$-matrix formulation \cite{MaWo} 
	of anti-self-dual Yang-Mills equations 
	and the quasi-Wronskian solution \cite{GHHN}.
	In subsection 4.1, we review exact one-soliton solutions  
	and the interpretation of soliton walls. 
	In subsection 4.2, we study the asymptotic behavior of 
	multi-soliton solution 
	and give the interpretation of intersecting soliton walls. 
	In section 5, we prove that the intersecting soliton walls 
	can be embedded into $G=\mathrm{SU}(2)$  gauge theory on the Ultrahyperbolic 
	space by showing that gauge fields are all anti-hermitian. 
	Section 6 is devoted to conclusion and discussion.

\section{Brief Introduction to Quasideterminants}

In this section, we give a brief introduction to  
quasideterminants defined firstly by Gelfand and Retakh
\cite{GeRe}. For detailed discussion, see e.g. \cite{GGRW, Huang}.
Briefly speaking, the quasideterminant of a $n \times n$ matrix X is 
a noncommutative generalization of the ratio of the determinant of X to the determinant of a  $(n-1) \times (n-1)$ submatrix.   
Therefore, quasideterminant is related to the inverse matrix of $X$. 
Here we assume the existence of the invertible matrix $X$. 

Let $X=(x_{ij})$ be a $n\times n$ invertible matrix 
over a noncommutative ring and 
$Y=(y_{ij})$ be the inverse matrix of $X$: $X Y=Y X =1$. 
Then the $(i,j)$-th quasideterminant of $X$ is defined 
as the inverse of an element of $Y=X^{-1}$:
\begin{eqnarray}
\vert X \vert_{ij}:=y_{ji}^{-1}.
\end{eqnarray}
A convenient representation for $(i,j)$-th quasideterminant is
\begin{eqnarray}
\label{Qdet}
\vert X\vert_{ij}=
\left|
\begin{array}{ccccc}
x_{11}&\cdots &x_{1j} & \cdots& x_{1n}\\
\vdots & & \vdots & & \vdots\\
x_{i1}&\cdots & {\fbox{$x_{ij}$}}& \cdots& x_{in}\\
\vdots & & \vdots & & \vdots\\
x_{n1}& \cdots & x_{nj}&\cdots & x_{nn}
\end{array}\right|.
\end{eqnarray}
To expand \eqref{Qdet},
let us introduce the inverse matrix formula for $2\times 2$ block matrix:
\begin{eqnarray*}
	\left(
	\begin{array}{cc}
		A&B \\C&d
	\end{array}
	\right)^{-1}
	=\left(\begin{array}{cc}
		A^{-1}+A^{-1} B S^{-1} C  A^{-1}
		&-A^{-1} B S^{-1}\\
		-S^{-1} C A^{-1}
		&S^{-1}
	\end{array}\right),
\end{eqnarray*}
where $A$ is a square matrix, $d$ is a single element and 
$S:=d-C A^{-1} B$ is called the Schur complement. 
We note that any invertible matrix 
can be decomposed into a $2\times 2$ block matrix 
and one of the diagonal terms is size $1 \times 1$.
We can assign $A$ to be $X^{ij}$, $B$ to be 
$X^{i}_{~j}$, $C$ to be $X_{i}^{~j}$, 
and $d$ to be $x_{ij}$, where $X^{ij}$ denotes the submatrix obtained from $X$ 
by deleting $i$-th row and $j$-th column, $X^{i}_{~j}$ and $X_{i}^{~j}$ denote the submatrices obtained from $j$-th column and $i$-th row of $X$ by deleting $x_{ij}$, respectively. Now
the $(i,j)$-th quasideterminant can be expressed as the Schur complement:
\begin{eqnarray}
\label{Schur complement}
\vert X \vert_{ij} 
&=&S = x_{ij} -  X_{i}^{~j} (X^{ij})^{-1} X^{i}_{~j}  \nonumber \\
&=&x_{ij}-\sum_{i^\prime (\neq i), j^\prime (\neq j)}
x_{ii^\prime}  (\vert {X}^{ij}\vert_{j^\prime i^\prime })^{-1}
x_{j^\prime j},
\end{eqnarray}
By using this, explicit representations of the 
quasideterminants can be obtained iteratively.
For example, for a $1\times 1$ matrix $X=x$
\begin{eqnarray*}
	\vert X \vert= x,
\end{eqnarray*}
and 
for a $2\times 2$ matrix $X=(x_{ij})$
\begin{eqnarray*}
	\vert X \vert_{11}=
	\begin{vmatrix}
		\fbox{$x_{11}$} &x_{12} \\x_{21}&x_{22}
	\end{vmatrix}
	=x_{11}-x_{12} x_{22}^{-1} x_{21},~~~
	\vert X \vert_{12}=
	\begin{vmatrix}
		x_{11} & \fbox{$x_{12}$} \\x_{21}&x_{22}
	\end{vmatrix}
	=x_{12}-x_{11} x_{21}^{-1} x_{22},\nonumber\\
	\vert X \vert_{21}=
	\begin{vmatrix}
		x_{11} &x_{12} \\ \fbox{$x_{21}$}&x_{22}
	\end{vmatrix}
	=x_{21}-x_{22} x_{12}^{-1} x_{11},~~~
	\vert X \vert_{22}=
	\begin{vmatrix}
		x_{11} & x_{12} \\x_{21}&\fbox{$x_{22}$}
	\end{vmatrix}
	=x_{22}-x_{21} x_{11}^{-1} x_{12}, 
\end{eqnarray*}
and for a $3\times 3$ matrix $X=(x_{ij})$
\begin{eqnarray*}
	\vert X \vert_{11}
	&=&
	\begin{vmatrix}
		\fbox{$x_{11}$} &x_{12} &x_{13}\\ x_{21}&x_{22}&x_{23}\\x_{31}&x_{32}&x_{33}
	\end{vmatrix}
	=x_{11}-(x_{12}, x_{13}) \left(
	\begin{array}{cc}x_{22} & x_{23} \\x_{32}&x_{33}\end{array}\right)^{-1}
	\left(
	\begin{array}{c}x_{21} \\x_{31}\end{array}
	\right)
	\nonumber\\
	&=&x_{11}-x_{12}  \begin{vmatrix}
		\fbox{$x_{22}$} & x_{23} \\x_{32}&x_{33}
	\end{vmatrix}^{-1}   x_{21}
	-x_{12} \begin{vmatrix}
		x_{22} & x_{23} \\\fbox{$x_{32}$}&x_{33}
	\end{vmatrix}^{-1}  x_{31}      \nonumber\\
	&&~~~~    -x_{13} \begin{vmatrix}
		x_{22} & \fbox{$x_{23}$} \\x_{32}&x_{33}
	\end{vmatrix}^{-1}  x_{21}
	-x_{13} \begin{vmatrix}
		x_{22} & x_{23} \\x_{32}&\fbox{$x_{33}$}
	\end{vmatrix}^{-1}  x_{31},
\end{eqnarray*}
and so on. We remark that the following expressions of quasideterminants are exactly the same because the Schur complement are all equivalent. 
\begin{eqnarray}
\label{schur}
\left|
\begin{array}{cc}
A&B \\C&\fbox{$d$}
\end{array}
\right|
=
\left|
\begin{array}{cc}
C&\fbox{$d$}\\A&B
\end{array}
\right|
=
\left|
\begin{array}{cc}
\fbox{$d$}&C \\B&A
\end{array}
\right|
=
\left|
\begin{array}{cc}
B&A \\\fbox{$d$}&C
\end{array}
\right|=d-CA^{-1}B.
\end{eqnarray}

Now let us introduce some important properties and identities of 
quasideterminants, which are relevant to discussions in this paper.

\vspace{2mm}
\noindent
{\bf Proposition 2.1 \cite{GGRW, GeRe, Huang}}  \\
Let $A=(a_{ij})$ be a square matrix of order $n$ in (i) $\sim$ (iii), 
while in (iv) and (v), appropriate partitions are made 
so that all matrices in quasideterminants are square. 
\begin{enumerate}
	\item [{(}i{)}]  Permutation of Rows and Columns
	
	The quasideterminant $\vert A\vert_{ij}$
	does not depend on permutations of rows and columns
	in the matrix $A$. 
	
	\item [{(}ii{)}]  The common multiplication of rows and columns
	
	For any invertible elements $\Lambda_{j}~(j=1,\cdots,n)$, 
	we have
	\begin{eqnarray}	
	\label{Rmulti}
	\left|
	\begin{array}{ccccc}
	a_{1,1}\Lambda_{1} & \cdots & a_{1,j}\Lambda_{j} & \cdots & a_{1,n}\Lambda_{n} \\
	\vdots &   & \vdots &  & \vdots  \\
	a_{i,1}\Lambda_{1} & \cdots & \fbox{$a_{i,j}\Lambda_{j}$} & \cdots & a_{i,n}\Lambda_{n} \\
	\vdots &   & \vdots &  & \vdots  \\
	a_{n,1}\Lambda_{1} & \cdots & a_{n,j}\Lambda_{j} & \cdots & a_{n,n}\Lambda_{n}
	\end{array}\right|
	=
	\left|
	\begin{array}{ccccc}
	a_{1,1} & \cdots & a_{1,j} & \cdots & a_{1,n} \\
	\vdots &   & \vdots &  & \vdots  \\
	a_{i,1} & \cdots & \fbox{$a_{i,j}$} & \cdots & a_{i,n} \\
	\vdots &   & \vdots &   & \vdots  \\
	a_{n,1} & \cdots & a_{n,j} & \cdots & a_{n,n}
	\end{array}\right|\Lambda_{j}
	\end{eqnarray}
	We note that in the left hand side of \eqref{Rmulti}, the common elements $\Lambda_{j}$ must appear 
	in the right side of the same column. 
	On the other hand, if the common elements appear in the left side of the same row, one can get the rule for common multiplication of rows.
	
	\item [{(}iii{)}] The addition of rows and columns
	
	Let the matrix $N=(n_{ij})$ be obtained from
	the matrix $A$ by replacing the $k$-th column of $A$
	with the sum of the $k$-th column and $l$-th column, that is,
	$n_{ik}=a_{ik}+a_{{il}}$ and $n_{ij}=a_{ij}$
	for $k\neq j$. Then
	\vspace{-2mm}
	\begin{eqnarray}
	\vert A\vert_{ij}=\vert N \vert_{ij},
	~~~\mbox{for}~j\neq k.
	\end{eqnarray}
	(The addition of rows is similar.) 
	
	\item [{(}iv{)}]  Noncommutative Jacobi identity \cite{GiNi07} 
	(An useful and simplified version of the noncommutative Sylvester's Theorem\cite{GeRe}):
	\begin{equation}
	\label{jacobi}
	\begin{vmatrix}
	a&R&b\\
	P&M&Q\\
	c&S&\fbox{$d$}
	\end{vmatrix}=
	\begin{vmatrix}
	M&Q\\
	S&\fbox{$d$}
	\end{vmatrix}-
	\begin{vmatrix}
	P&M\\
	\fbox{$c$}&S
	\end{vmatrix}    
	\begin{vmatrix}
	\fbox{$a$}&R\\
	P&M
	\end{vmatrix}^{-1}
	\begin{vmatrix}
	R&\fbox{$b$}\\
	M&Q
	\end{vmatrix}.
	\end{equation}
	
	\item [{(}v{)}] 
	Homological relations \cite{GeRe,GiNi07}
	\begin{eqnarray}
	\!\!    \begin{vmatrix}
	a&\!R\!&b\\
	P&\!M\!&Q\\
	\fbox{$c$}&\!S\!&d
	\end{vmatrix}
	\!=\!  \begin{vmatrix}
	a&\!R\!&b\\
	P&\!M\!&Q\\
	c&\!S\!&\fbox{$d$}
	\end{vmatrix}\!
	\begin{vmatrix}
	a&\!R\!&b\\
	P&\!M\!&Q\\
	\fbox{0}&\!0\!&1
	\end{vmatrix},~
	\label{homological}
	\begin{vmatrix}
	a&\!R\!&\fbox{$b$}\\
	P&\!M\!&Q\\
	c&\!S\!&d
	\end{vmatrix}
	\!=\!      \begin{vmatrix}
	a&\!R\!&\fbox{0}\\
	P&\!M\!&0\\
	c&\!S\!&1
	\end{vmatrix}\!
	\begin{vmatrix}
	a&\!R\!&b\\
	P&\!M\!&Q\\
	c&\!S\!&\fbox{$d$}
	\end{vmatrix}
	\end{eqnarray}
	 \mbox{If we use the homological relation again on the right hand side, we can obtain the}\\ \mbox{following inverse relation immediately:}
		\begin{eqnarray} \label{inverse relation}
		\begin{vmatrix}
		a&\!R\!&b\\
		P&\!M\!&Q\\
		\fbox{0}&\!0\!&1
		\end{vmatrix}^{-1}
		=
		\begin{vmatrix}
		a&\!R\!&b\\
		P&\!M\!&Q\\
		1&\!0\!&\fbox{0}
		\end{vmatrix},~~
		\begin{vmatrix}
		a&\!R\!&\fbox{0}\\
		P&\!M\!&0\\
		c&\!S\!&1
		\end{vmatrix}^{-1}\!
		=
		\begin{vmatrix}
		a&\!R\!&1\\
		P&\!M\!&0\\
		c&\!S\!&\fbox{0}
		\end{vmatrix}\!
		\end{eqnarray}
\end{enumerate}

We note that the definition of the quasideterminants 
and Proposition 2.1 are valid even if the matrix elements belong to
noncommutative associative algebras. 
This means that we can consider all elements $x_{ij}$ in \eqref{Qdet}
as $N\times N$ matrices. (In this case, $X$ is $nN \times nN$.) For our purpose in this paper, we will just consider the case of  $2 \times 2$ matrices because the gauge group is $G=GL(2, \mathbb{C})$.
In this case, we can expand the quasideterminant as a $2 \times 2$ matrix and the matrix elements are four quasideterminants.
\begin{eqnarray}
\label{2x2}
\begin{vmatrix}
M\!\!\!\!\!&\begin{array}{cc}C_1&C_2\end{array}\\
\begin{array}{c}
R_1\!\!\!\!\!\\R_2\!\!\!\!\!
\end{array}
&\fbox{$
	\begin{array}{cc}
	a&b\\c&d
	\end{array}$}
\end{vmatrix}&
=
\left(
\begin{array}{cc}
\begin{vmatrix}
M&C_1\\
R_1&\fbox{$a$}
\end{vmatrix}&
\begin{vmatrix}
M&C_2\\
R_1&\fbox{$b$}
\end{vmatrix}
\\
\begin{vmatrix}
M&C_1\\
R_2&\fbox{$c$}\\
\end{vmatrix}&
\begin{vmatrix}
M&C_2\\
R_2&\fbox{$d$}\\
\end{vmatrix}
\end{array}
\right).
\end{eqnarray}
Note that in the right hand side, each box includes a $1\times 1$ scalar element. 
Hence they can be represented as a ratio of ordinary determinants 
by virtue of the Laplace formula on inverse matrices:
\begin{eqnarray}
\label{laplace}
\left|
X
\right|_{ij}
=y_{ji}^{-1}
=(-1)^{i+j}\frac{\det X}{\det X^{ij}},
\end{eqnarray}
where $X^{ij}$ is a matrix obtained from $X$ 
by deleting $i$-th row and $j$-th column. 

\vspace{3mm}

\section{ASDYM Equations and quasi-Wronskian Solutions}

In this section, we review anti-self-dual Yang-Mills equations in 
four-dimensional flat spaces 
whose real coordinates are denoted by $x^\mu~(\mu=0,1,2,3)$. 
To facilitate the discussion, we set the gauge group 
to be $G=GL(N, \mathbb{C})$ (or subgroup of $GL(N,\mathbb{C})$). 

Firstly, let us consider a four-dimensional complex space 
with coordinates $(z,\widetilde z, w, \widetilde w)$ 
and define the metric to be $ds^2=2(dz d\widetilde z -dw d\widetilde w)$. 
We can recover various real spaces from this complex space 
by imposing suitable conditions on $z,\widetilde z, w, \widetilde w$. 
We call them the real slice conditions in this paper. 
For example, the Euclidean real space $\mathbb{E}$ is given by 
$\widetilde z=\overline z,~
\widetilde w= -\overline w$, 
the Minkowski real space $\mathbb{M}$ is by $z, 
\widetilde z \in \mathbb{R},~\widetilde w= \overline w$, 
and the Ultrahyperbolic real space $\mathbb{U}$ is by
$z, \widetilde z,w, \widetilde w\in \mathbb{R}$. 
Explicit relations between 
$z,\widetilde z, w, \widetilde w$ and  $x^\mu$ 
are summarized in Table \ref{table} in subsection 4.1. 

Now we introduce an equivalent representation 
of the anti-self-dual Yang-Mills equations on this complex space,
called Yang's equation :
\begin{eqnarray}
\label{yang}
\partial_{\widetilde{z}}(\partial_z J \cdot J^{-1})
-\partial_{\widetilde{w}} (\partial_w J \cdot J^{-1} )=0,
\end{eqnarray}
where $J$ is an $N\times N$ complex matrix. 
This formulation gives a more concise way to unify 
anti-self-dual Yang-Mills equations of various real spaces 
to one complex form and extensively used 
in the field of integrable systems. Moreover, 
the anti-self-dual gauge fields can be expressed explicitly 
by the solution $J$ of Yang's equation in a convenient gauge:
\begin{eqnarray}
\label{gauge_f_special}
A_{z}=-\partial_{z}J\cdot J^{-1},~~A_{w}=-\partial_{w}J\cdot J^{-1},~~
A_{\widetilde z} = A_{\widetilde w} = 0,
\end{eqnarray}
We can easily to check that \eqref{gauge_f_special}
actually satisfies the anti-self-dual 
Yang-Mills equation of a complex representation:
\begin{eqnarray}
F_{zw}=0,~~~
F_{\widetilde{z}\widetilde{w}}=0,~~~
F_{z\widetilde{z}}-F_{w\widetilde{w}}=0,
\label{asdym}
\end{eqnarray}
where $F_{zw}:=\partial_z A_w-\partial_w A_z+[A_z,A_w]$ and so on 
denote the field strengths. 
By taking real slice conditions on $z, \widetilde{z}, w, \widetilde{w}$
as mentioned above, \eqref{asdym}
reduces to the standard anti-self-dual Yang-Mills 
equations in four-dimensional real spaces 
in the sense of the Hodge dual operator.

For the sake of generating exact solutions systematically, 
a typical technique developed in the field of integrable systems 
is to find the Lax representations, 
and then use the covariance of Lax equations 
under the Darboux transformation 
to generate more exact solutions. 
Here we introduce a formulation, 
slightly different from the conventional Lax formalism. 
Let us consider the following linear system \cite{GNO00}:
\begin{eqnarray}
L(\phi)&:=&
J \partial_{w}(J^{-1} \phi)
- (\partial_{\widetilde{z}}\phi)\zeta=0,\nonumber\\
M(\phi)&:=&
J \partial_{z}(J^{-1} \phi)
- (\partial_{\widetilde{w}}\phi)\zeta=0,
\label{lin_yang}
\end{eqnarray}
where $\zeta$ is 
a matrix generalization of the spectral parameter.
More precisely, $\zeta$ is an $N\times N$ constant matrix.  
We can show that the compatibility condition 
$L(M(\phi))-M(L(\phi))=0$ 
implies Yang's equation \eqref{yang} and  
the linear system \eqref{lin_yang} is covariant under
the following Darboux transformation \cite{GNO00}:
\begin{eqnarray}
\label{Darboux_phi}
\widetilde{\phi}=
\phi \zeta - \theta \Lambda \theta^{-1} \phi,~~~
\label{Darboux_J}
\widetilde{J}= -\theta \Lambda \theta^{-1} J,
\end{eqnarray}
where $\theta$ is an eigenfunction of the linear system
\eqref{lin_yang}
for the choice of eigenvalue $\zeta=\Lambda$. 

After $n$ iteration of the Darboux transformation, 
we get an exact solution $J_n$ 
of Yang's equation from a trivial seed solution $J=1$
and this kind of solution can be expressed 
in terms of quasideterminant in a compact form \cite{GNO00}. 
Here we use the terminology Wronskian type quasideterminants, 
or quasi-Wronskian for short.
The specific form of $J_{n}$ is as follows : 
\begin{eqnarray}
\label{Jn}
J_{n}=
\left|
\begin{array}{cccc}
\theta_1&\cdots&\theta_n& 1\\
\theta_1\Lambda_1&\cdots &\theta_n\Lambda_n& 0\\
\vdots   && \vdots& \vdots\\
\theta_1\Lambda_1^{n-1}&\cdots&\theta_n\Lambda_n^{n-1}& 0\\
\theta_1\Lambda_1^{n}&\cdots& \theta_n\Lambda_n^{n}& \fbox{$0$}
\end{array}\right|,
\end{eqnarray}
where $(\theta_i, \Lambda_i)~(i=1,2,\cdots,n)$ are pairs of 
eigenfunctions and eigenvalues of the 
initial linear system for $J=1$ in \eqref{lin_yang} \cite{GHHN,GNO00}:
\begin{eqnarray}
\label{chasing}
\partial_w \theta_i=(\partial_{\widetilde{z}}\theta_i)\Lambda_i,~~~
\partial_z \theta_i=(\partial_{\widetilde{w}}\theta_i)\Lambda_i.
\end{eqnarray} 
We remark that in $G=GL(N, \mathbb{C})$ case, all the $\theta_i$ and $\Lambda_i$ are $N \times N$ matrices. Therefore, $J_n$ is exactly a $N \times N$ matrix if we expand it term by term as the Schur complement \eqref{Schur complement}.

\section{Multi-Soliton Solutions of ASDYM equation}

As mentioned in previous section,
we can obtain  $n$ different pairs of 
$(\theta_i, \Lambda_i)$ by solving \eqref{chasing} and 
use them to form exact solutions $J_n$  of Yang's equation. 
One kind of $n$-soliton solutions for $G=\mathrm{GL(2,\mathbb{C})}$ 
is given by an interesting case of $J_n$ 
which is composed of a special set of solutions of \eqref{chasing} 
\cite{GHHN}:
\begin{eqnarray}
\label{CS_n}
\theta_i=\left(
\begin{array}{cc}
a_i e^{L_i}
& 
b_i e^{-M_i}
\\ 
c_i e^{-L_i}
& 
d_i e^{M_i}
\end{array}\right)
,~~~
\Lambda_i=
\left(
\begin{array}{cc}
\lambda_i & 0 \\
0 & \mu_i
\end{array}
\right),
\end{eqnarray}
where $i=1,2,\cdots,n$,  
$L_i
=\lambda_i \beta_i z
+\alpha_i\widetilde{z}
+\lambda_i \alpha_i w
+\beta_i\widetilde{w},~
M_i:=\mu_i \delta_i z
+\gamma_i\widetilde{z}
+\mu_i\gamma_i w
+\delta_i\widetilde{w}$, 
and 
$\lambda_i, \mu_i, a_i, b_i, c_i, d_i, 
\alpha_i,\beta_i,\gamma_i, \delta_i$ are complex constants. 
Furthermore, we confirm that 
in subsection 4.2, a reduced version of  solution \eqref{CS_n} (Cf. \eqref{U2_n})
actually generates a multi-soliton distribution 
because this distribution 
has $n$ localized action densities in the asymptotic region.
Hence this multi-soliton is stable in the scattering process.

\subsection{One-Soliton Solutions of ASDYM equation}

In this subsection we summarize some results and properties
of the one-soliton solutions in our previous work \cite{HaHu}. 
The $J$-matrix is
\begin{eqnarray}
\label{1soliton_cpx}
J=
\left|
\begin{array}{cc}
\theta & 1  \\
\theta \Lambda & \ \fbox{0}
\end{array}
\right|
=
-\theta \Lambda \theta^{-1},~~~
\theta=\left(
\begin{array}{cc}
ae^{L}
& 
be^{-M}
\\ 
ce^{-L}
& 
de^{M} 
\end{array}\right),~~~
\Lambda=\left(
\begin{array}{cc}
\lambda
& 
0
\\ 
0
& 
\mu 
\end{array}\right).
\end{eqnarray}
The action density of this kind of solution is calculated 
as follows: 
\begin{eqnarray}
\label{action_cpx_1}
{\mbox{Tr}} F^2&=&
8(\lambda-\mu)^2(\alpha\delta-\beta\gamma)^2
\left(
2{\mbox{sech}}^2 X-3{\mbox{sech}}^4 X
\right),\\	
&&\mbox{where } ~X:=M+L+\dfrac12 \log(-ad/bc).
\label{X} 
\end{eqnarray}
We remark that \eqref{action_cpx_1} is a highly nontrivial result 
and cannot be realized in the conventional Lax formalism. 
More explicitly, if the spectral parameter matrix $\Lambda$ is 
a scalar matrix (i.e. $\lambda=\mu$), then linear system \eqref{lin_yang} 
is equivalent to the conventional one and 
the resulting action density 
\eqref{action_cpx_1} becomes trivial : $\mbox{Tr}F^2=0$.

By imposing real slice conditions on the space-time coordinates 
$z, \widetilde{z}, w, \widetilde{w}$, 
and putting an additional condition on $J$-matrix: 
$d=\overline{a}, c=-\overline{b}$
and $M=\overline{L}$, we can get one-soliton solutions on 
different real spaces $\mathbb{E}$, $\mathbb{M}$ and  $\mathbb{U}$:
\begin{eqnarray}
\label{U2_1}
J=
\left|
\begin{array}{cc}
\theta & 1  \\
\theta \Lambda & \ \fbox{0}
\end{array}
\right|
=
-\theta \Lambda \theta^{-1},~~~
\theta=\left(
\begin{array}{cc}
a e^{L}
& 
b e^{-\overline{L}}
\\ 
-\overline{b} e^{-L}
& 
\overline{a} e^{\overline{L}}
\end{array}\right)
,~~~
\Lambda=
\left(
\begin{array}{cc}
\lambda & 0 \\
0 & \mu
\end{array}
\right).
\end{eqnarray}
The resulting action densities take real value:
\begin{eqnarray}
\label{action_cpx_2}
{\mbox{Tr}} F_{\mu\nu}F^{\mu\nu}=C\left(
2{\mbox{sech}}^2 X-3{\mbox{sech}}^4 X
\right),	
\end{eqnarray}
where 
$X:=L+\overline{L}+\log\vert a/b\vert$
and $C$ is a real constant depending on the signature 
of different real spaces. 
We note that the principal peak of these action densities lie 
on a three-dimensional hyperplane 
defined by $X=L+ \overline{L}+\log|a /b |=0$
with normal vector $l_\mu+\overline{l}_\mu$.  
Therefore, we can interpret them as codimension-one solitons 
in four dimensional spaces and use the terminology 
soliton walls in this paper to distinguish our solution from domain walls.

Main results of \cite{HaHu} are summarized in Table \ref{table}:

\vspace{-8mm}
\noindent
\begin{table}[h]
	\caption{Summary of One-Soliton Solutions}
	\label{table}
	\bigskip
	\begin{tabular}{|c|c|c|c|}
		\hline
		signature & $\mathbb{E}$  & $\mathbb{M}$ & $\mathbb{U}$  \\\hline\hline
		real slice & $\widetilde z=\overline z,~
		\widetilde w= -\overline w$  & 
		$z, \widetilde z \in \mathbb{R},~\widetilde w= \overline w$ & 
		$z, \widetilde z,w, \widetilde w\in \mathbb{R}$
		\\
		$\displaystyle
		\sqrt{2}
		\left(\begin{array}{cc}\!\!\widetilde{z}&w\!\!\\\!\!\widetilde{w}&z\!\!\end{array}\right)
		\!\!=\!\!$
		&
		$\!\!\displaystyle
		\left(\begin{array}{cc}\!\!x^0+ix^1\!\!&-x^2+ix^3\!\!\\\!\!x^2+ix^3\!\!&
		x^0-ix^1\!\!
		\end{array}\right)\!\!$
		&
		$\displaystyle
		\left(\begin{array}{cc}\!\!x^0+x^1&x^2-ix^3\!\!\\\!\!x^2+ix^3&x^0-x^1\!\!
		\end{array}\right)$
		&
		$\displaystyle \!\!
		\left(\begin{array}{cc}\!\!x^0+x^2&x^1-x^3\!\!\\\!\!x^1+x^3&x^0-x^2\!\!
		\end{array}\right)\!\!$
		\\\hline
		$\!\!$reality condition$\!\!$&$\mu=-1/\overline{\lambda}$&NONE&$\mu=\overline{\lambda}$
		\\
		$L=\overline{M}$ & $L=(\lambda \beta) z
		+\alpha\overline{z}$  & 
		$L=(\lambda \overline{\mu}\alpha) z
		+\alpha\widetilde{z}$
		& 
		$L=(\lambda \beta) z
		+\alpha\widetilde{z}
		$ \\
		&$+(\lambda \alpha) w
		-\beta \overline{w}$
		&$+(\lambda \alpha) w
		+(\overline{\mu}\alpha)\overline{w} $
		&$+(\lambda \alpha) w
		+\beta \widetilde{w}$
		\\\hline
		$L=l_\mu x^\mu$ & 
		$l_\mu\!\!=\!\!\displaystyle\frac{1}{\sqrt{2}}\left(\begin{array}{c}
		\!\!\alpha+\lambda\beta\!\!\\\!\!i(\alpha-\lambda\beta)\!\! \\\!\!\beta-\lambda\alpha\!\! \\\!\!i(\beta+\lambda\alpha)\!\!
		\end{array}\right)$ & 
		$l_\mu\!\!=\!\!\displaystyle\frac{\alpha}{\sqrt{2}}\left(\begin{array}{c}
		1+\lambda \overline{\mu} \\ 1-\lambda \overline{\mu}\\\overline{\mu}+\lambda \\i(\overline{\mu}-\lambda)
		\end{array}\right)$  & 
		$l_\mu\!\!=\!\!\displaystyle\frac{1}{\sqrt{2}}\left(\begin{array}{c}
		\!\!\alpha+\lambda\beta\!\!\\\!\!\beta-\lambda\alpha\!\! \\\!\!\alpha-\lambda\beta\!\! \\\!\!\beta+\lambda\alpha\!\!
		\end{array}\right)$  \\\hline
		constant $C$&$\!\!8(\left|\alpha\right|^2\!\!+\!\left|\beta\right|^2)^2
		(\left|\lambda \right|^2+1)^2\!\!$&
		$8\left|{\alpha}^2(\lambda-\mu)\right|^{4}$&
		$\!\!8\left[(\alpha\overline{\beta}\!-\!\overline{\alpha}\beta)
		(\lambda \!-\!\overline{\lambda})  \right]^2\!\!$\\\hline
		hermicity&$A_0,A_2$:anti-hermitian&$\!\!A_0,A_1,A_2$:anti-hermitian$\!\!$&
		$A_0,A_1,A_2,A_3$:\\
		of $A_\mu$&$A_1,A_3$:hermitian&$A_3$:hermitian&anti-hermitian\\
		&(when $\lambda=\pm i$)&(when $\lambda=\overline{\mu}$)&(NONE)
\\\hline
		gauge group&$\mathrm{G=SL(2,\mathbb{C})}$&$\mathrm{G=SL(2,\mathbb{C})}$&$\mathrm{G=SU(2)}$\\\hline
\end{tabular} 
\end{table}

\subsection{Asymptotic Behavior of the $n$-Soliton Solutions}

Now let us put the condition: $d_{i}=\overline{a}_{i},~ c_{i}=-\overline{b}_{i},~  M_{i}=\overline{L}_{i}$ on \eqref{CS_n} and discuss the asymptotic behavior of 
the $n$-soliton solution ($i=1,2,\cdots,n$):
\begin{eqnarray}
\label{U2_n}
J_{n}=
\left|
\begin{array}{cccc}
\theta_1&\cdots&\theta_n& 1\\
\theta_1\Lambda_1&\cdots &\theta_n\Lambda_n& 0\\
\vdots   && \vdots& \vdots\\
\theta_1\Lambda_1^{n}&\cdots& \theta_n\Lambda_n^{n}& \fbox{$0$}
\end{array}\right|, ~ 
\theta_i=\left(
\begin{array}{cc}
a_i e^{L_i}
& 
b_i e^{-\overline{L}_i}
 \\ 
-\overline{b}_i e^{-L_i}
  & 
\overline{a}_i e^{\overline{L}_i}
\end{array}\right)
,~
\Lambda_i=
\left(
\begin{array}{cc}
\lambda_i & 0 \\

0 & \mu_i
\end{array}
\right),
\end{eqnarray}
where
$L_i
=\lambda_i \beta_i z
+\alpha_i\widetilde{z}
+\lambda_i \alpha_i w
+\beta_i\widetilde{w}
$, 
and 
$a_i, b_i, \alpha_i,\beta_i, \lambda_i,\mu_i\in \mathbb{C}$. 
We assume here that $L_i$ $(i=1,2,\cdots,n)$ are independent 
with each other and there is no special relation between them.  
In other words, we just consider the situation of pure-soliton scattering 
and exclude the case of resonance processes.

Inspired by a typical technique that was used for discussing 
the asymptotic behavior of Wronskian type $n$-soliton solutions 
of the KP equation, 
we follow a quite similar procedure like this to deal with 
quasi-Wronskian type $n$-soliton solutions 
of the anti-self-dual Yang-Mills equations. 
It is actually a new attempt 
because the elements $\theta_{i}, \Lambda_{i}$ in the quasi-Wronskian  
$J_{n}$ are $2\times 2$ matrices rather than a scalar function. (Cf. \eqref{2x2})
If the reader prefer the Wronskian determinants, 
just consider each matrix elements in the $2\times 2$ matrix $J_{n}$ 
as the ratio of the Wronskian determinants (Cf. \eqref{laplace}).
If doing so, you will find that the analysis becomes much more complicated than the discussion of using quasideterminants.

Firstly, let us pick an $I\in\left\{1, 2, ..., n\right\}$ 
and keep $L_I$ (and $\overline{L}_I$) to be finite. 
(This is in fact equivalent to consider a comoving frame with 
the $I$-th soliton.)
If we take the asymptotic limit $r:=((x^0)^2+(x^1)^2+(x^2)^2+(x^3)^2)^{1/2}
\rightarrow \infty$ on $J_{n}$.  
Then for $i\neq I$, Re$ L_i$ ($=$ Re$\overline{L}_i$) 
goes to (i) positive infinity or (ii) negative infinity, that is, 
$\vert e^{L_i} \vert $ $(=\vert e^{\overline{L}_i}\vert)$ 
goes to (i) positive infinity or (ii) zero. 
Now we can use \eqref{Rmulti} to eliminate common factors in each column ($i \neq I$) of $J_n$ and take $r \rightarrow \infty$ to obtain the following asymptotic form:
\begin{eqnarray*}
J_{n} \rightarrow 
\left|
\begin{array}{cccccc}
C_1
&
\cdots
&
\theta_I
& 
\cdots
&
C_n
& 
1
\\
C_1 \Lambda_1
&
\cdots
&
\theta_I\Lambda_I
&
\cdots
& 
C_n\Lambda_n
&0
\\
\vdots
&
&
\vdots
&
& 
\vdots
&
\vdots
\\
C_1\Lambda_1^n
& 
\cdots
&
\theta_I\Lambda_I^n
& 
\cdots
&
C_n\Lambda_n^n
& \fbox{0}
\end{array}
\right|
~~~
\mbox{where (i)}
~
 C_i=\left(
\begin{array}{cc}
1&0 \\
0&1 
\end{array}
\right)
~\mbox{or (ii)}~
\left(
\begin{array}{cc}
0&1 \\
-1&0 
\end{array}
\right).
\end{eqnarray*}
Next, our goal is to remove all the $C_{i}$ 
from the asymptotic form such that 
the remaining elements  
are just $I$-th column and constant matrices $\Lambda_{i}, 
\mbox{for} ~i \neq I$. 
The case (i) is trivial, while the case (ii) can be done by adjusting the diagonal terms of $\Lambda_i$ to get the commutation relation:
\begin{eqnarray}
\label{flip}
C_i \Lambda_i^k = 
\left(
\begin{array}{cc}
0&1 \\
-1&0 
\end{array}
\right)
\left(
\begin{array}{cc}
\lambda_i^k&0 \\
0&\mu_i^k
\end{array}
\right)
=
\left(
\begin{array}{cc}
\mu_i^k&0 \\
0&\lambda_i^k 
\end{array}
\right)
\left(
\begin{array}{cc}
0&1 \\
-1&0 
\end{array}
\right),
\end{eqnarray}
and by using \eqref{Rmulti} to remove all the right common factors. 
The explicit result is
\begin{eqnarray*}
	&&J_{n} \rightarrow 
	\left|
	\begin{array}{cccccc}
		1
		&
		\cdots
		&
		\theta_I
		& 
		\cdots
		&
		1
		& 
		1
		\\
		\Lambda_1^{(\pm)}
		&
		\cdots
		&
		\theta_I\Lambda_I
		&
		\cdots
		& 
		\Lambda_n^{(\pm)}
		&0
		\\
		\vdots
		&
		&
		\vdots
		&
		& 
		\vdots
		&
		\vdots
		\\
		\Lambda_1^{(\pm)n}
		& 
		\cdots
		&
		\theta_I\Lambda_I^n
		& 
		\cdots
		&
		\Lambda_n^{(\pm)n}
		& \fbox{0}
	\end{array}
	\right|,  \\
&& \Lambda_{i,~ i \neq I}^{(\pm)}:=
\left(
\begin{array}{cc}
	\lambda_i^{(\pm)}&0 \\
	0&\lambda_i^{(\mp)}
\end{array}
\right)
=
\left\{
\begin{array}{ll}
	\left(
	\begin{array}{cc}
		\lambda_i^{(+)}&0 \\
		0&\lambda_i^{(-)}
	\end{array}
	\right)
	& \mbox{(i) Re} L_i\rightarrow +\infty ~~(\vert e^{L_i}\vert \rightarrow \infty)\\
	\left(
	\begin{array}{cc}
		\lambda_i^{(-)}&0 \\
		0&\lambda_i^{(+)}
	\end{array}
	\right)
	& \mbox{(ii) Re} L_i\rightarrow -\infty~~(\vert e^{L_i}\vert \rightarrow 0)
\end{array}
\right.
\end{eqnarray*}
where we introduce new notations: $\lambda_i^{(+)}:=\lambda_i,~ \lambda_i^{(-)}:=\mu_i$ to unify various cases of (i) and (ii) in the discussions and proofs of this subsection.

Now we need to to show that the asymptotic form of $J_n$ is essentially the same as one-soliton solution. (Cf. \eqref{U2_1} with $\theta=\theta_I$.)
Without loss of generality, we can 
consider the case of $I=1$ for convenience. 
The discussions of other cases are the same 
because of the permutation property of 
the quasideterminants (Propositon 2.1 (i)).
By applying the noncommutative Jacobi identity \eqref{jacobi}
and \eqref{Rmulti} 
to the asymptotic form, 
$J_n$ can be represented  as the product of 
three kind of quasideterminants defined as follows:
\begin{eqnarray}
\label{asymp_Jn}
J_{n} &\rightarrow&
\left|
\begin{array}{ccccc}
\theta_1
& 
1
&
\cdots
&
1
& 
1
\\
\theta_1\Lambda_1
&
\Lambda_2^{(\pm)}
&
\cdots
& 
\Lambda_n^{(\pm)}
&
0
\\
\vdots
&
\vdots
&
&
\vdots
&
\vdots
\\
\theta_1\Lambda_1^{n-1}
& 
\Lambda_2^{(\pm)n-1}
& 
\cdots
& 
\Lambda_{n}^{(\pm)n-1}
&0
\\
\theta_1
\Lambda_1^{n}
& 
\Lambda_2^{(\pm)n}
& 
\cdots
& 
\Lambda_{n}^{(\pm)n}
& \fbox{0}
\end{array}
\right|
=-Q_{n} \Lambda_1 Q_n^{-1} D_{n},
\end{eqnarray}
where
\begin{eqnarray*}
Q_n:=
\left|
\begin{array}{cccc}
\theta_1
& 
1
&
\cdots
& 
1
\\
\theta_1\Lambda_1
&
\Lambda_2^{(\pm)}
&
\cdots
& 
\Lambda_n^{(\pm)}
\\
\vdots
&
\vdots
&
&
\vdots
\\
\theta_1\Lambda_1^{n-2}
& 
\Lambda_2^{(\pm)n-2}
& 
\cdots
& 
\Lambda_{n}^{(\pm)n-2}
\\
\fbox{$\theta_1\Lambda_1^{n-1}$}
& 
\Lambda_2^{(\pm)n-1}
& 
\cdots
& 
\Lambda_{n}^{(\pm)n-1}
\end{array}
\right|,~~~
D_n:=
\left|
\begin{array}{cccc}
\fbox{0}
&
1
&
\cdots
& 
1
\\
0
&
\Lambda_2^{(\pm)}
&
\cdots
& 
\Lambda_n^{(\pm)}
\\
\vdots
&
\vdots
&
&
\vdots
\\
0
&
\Lambda_2^{(\pm)n-2}
& 
\cdots
& 
\Lambda_{n}^{(\pm)n-2}
\\
1
&
\Lambda_2^{(\pm)n-1}
& 
\cdots
& 
\Lambda_{n}^{(\pm)n-1}
\end{array}
\right|^{-1}.
\end{eqnarray*}
In fact, $Q_{n}$ and $D_{n}$ can be expanded explicitly as the following \eqref{Qn} and \eqref{Dn}, respectively.  The proof is made by using mathematical induction. 
For $n\geq 2$:
\begin{eqnarray}
\label{Qn}
Q_n\!\!&=&\!\!
\left(
\begin{array}{cc}
\!(\lambda_1-\lambda_2^{(\pm)})\cdots
(\lambda_1-\lambda_n^{(\pm)})a_1e^{L_1}
&
(\mu_1-\lambda_2^{(\pm)})\cdots
(\mu_1-\lambda_n^{(\pm)})b_1e^{-\overline{L}_1}\!
\\
\!-(\lambda_1-\lambda_2^{(\mp)})\cdots
(\lambda_1-\lambda_n^{(\mp)})\overline{b}_1e^{-L_1}
&
(\mu_1-\lambda_2^{(\mp)})\cdots
(\mu_1-\lambda_n^{(\mp)})a_1e^{\overline{L}_1}\!
\end{array}
\right)\!\!,~~~~~~\\
D_n\!\!&=&\!\!(-1)^{n-1} \Lambda_n^{(\pm)}\cdots \Lambda_2^{(\pm)}.
\label{Dn}
\end{eqnarray}
For $n=2$, the statement is clearly true because 
we can easily check that 
\begin{eqnarray*}
Q_{2}
 =
\left|
\begin{array}{cc}
\theta_{1} & 1  \\
\fbox{ $\theta_{1} \Lambda_{1}$} & \Lambda_{2}^{(\pm)}
\end{array}
\right|
=\theta_{1}\Lambda_{1}-\Lambda_2^{(\pm)}\theta_1,~~~
D_{2} =
		\left|
		\begin{array}{cc}
			\fbox{0} & 1  \\
			1 & \Lambda_{2}^{(\pm)}
		\end{array}
		\right|^{-1}
=	-\Lambda_2^{(\pm)}
\end{eqnarray*}
by the Schur complements \eqref{schur}.
Let us assume that the statement holds for some $n$. 
By using the noncommutative Jacobi identity \eqref{jacobi}
and \eqref{Rmulti}, we have 
\begin{eqnarray*}
&&
Q_{n+1}
=
\left|
\begin{array}{ccccc}
\theta_1\Lambda_1
& 
\Lambda_2^{(\pm)}
& 
\cdots
& 
\Lambda_n^{(\pm)}
\\
\theta_1\Lambda_1^{2}
&
\Lambda_2^{(\pm)2}
&
\cdots
& 
\Lambda_{n}^{(\pm)2}
\\
\vdots
&
\vdots
&
&
\vdots
\\
\!\!\fbox{$\theta_1 \Lambda_1^{n}$}\!\
&
\!\!\Lambda_2^{(\pm)n}\!\!
& 
\!\!\cdots\!\!
& 
\!\!\Lambda_{n}^{(\pm)n}\!\!
\end{array}
\right|
\\
&&
-
\left|
\begin{array}{ccc}
\Lambda_2^{(\pm)}
& 
\cdots
& 
\Lambda_{n+1}^{(\pm)}
\\
\Lambda_2^{(\pm)2}
&
\cdots
& 
\Lambda_{n+1}^{(\pm)2}
\\
\vdots
&
&
\vdots
\\
\!\!\Lambda_2^{(\pm)n}\!\!
& 
\!\!\cdots\!\!
& 
\!\!\fbox{$\Lambda_{n+1}^{(\pm)n}$}\!\!
\end{array}
\right|
\left|
\begin{array}{ccc}
1
& 
\cdots
& 
\fbox{$1$}
\\
\Lambda_2^{(\pm)}
&
\cdots
& 
\Lambda_{n+1}^{(\pm)}
\\
\vdots
&
&
\vdots
\\
\!\!\Lambda_2^{(\pm)n-1}\!\!
& 
\!\!\cdots\!\!
& 
\!\!\Lambda_{n+1}^{(\pm)n-1}\!\!
\end{array}
\right|^{-1}
\!\!
\left|
\begin{array}{ccccc}
\fbox{$\theta_1$}
& 
1
& 
\cdots
& 
1
\\
\theta_1\Lambda_1
&
\Lambda_2^{(\pm)}
&
\cdots
& 
\Lambda_{n}^{(\pm)}
\\
\vdots
&
\vdots
&
&
\vdots
\\
\!\!\theta_1 \Lambda_1^{n-1}\!\!
&
\!\!\Lambda_2^{(\pm)n-1}\!\!
& 
\!\!\cdots\!\!
& 
\!\!\Lambda_{n}^{(\pm)n-1}\!\!
\end{array}
\right|\\
&&
=
\left|
\begin{array}{ccccc}
\theta_1
& 
1
& 
\cdots
& 
1
\\
\theta_1\Lambda_1
&
\Lambda_2^{(\pm)}
&
\cdots
& 
\Lambda_{n}^{(\pm)}
\\
\vdots
&
\vdots
&
&
\vdots
\\
\!\!\fbox{$\theta_1 \Lambda_1^{n-1}$}\!\
&
\!\!\Lambda_2^{(\pm)n-1}\!\!
& 
\!\!\cdots\!\!
& 
\!\!\Lambda_{n}^{(\pm)n-1}\!\!
\end{array}
\right| \Lambda_1 
\\
&&
-
\left|
\begin{array}{ccc}
1
& 
\cdots
& 
1
\\
\Lambda_2^{(\pm)}
&
\cdots
& 
\Lambda_{n+1}^{(\pm)}
\\
\vdots
&
&
\vdots
\\
\!\!\Lambda_2^{(\pm)n-1}\!\!
& 
\!\!\cdots\!\!
& 
\!\!\fbox{$\Lambda_{n+1}^{(\pm)n-1}$}\!\!
\end{array}
\right|
\Lambda_{n+1}^{(\pm)}
\left|
\begin{array}{ccc}
1
& 
\cdots
& 
\fbox{$1$}
\\
\Lambda_2^{(\pm)}
&
\cdots
& 
\Lambda_{n+1}^{(\pm)}
\\
\vdots
&
&
\vdots
\\
\!\!\Lambda_2^{(\pm)n-1}\!\!
& 
\!\!\cdots\!\!
& 
\!\!\Lambda_{n+1}^{(\pm)n-1}\!\!
\end{array}
\right|^{-1}
\!\!
\left|
\begin{array}{ccccc}
\fbox{$\theta_1$}
& 
1
& 
\cdots
& 
1
\\
\theta_1\Lambda_1
&
\Lambda_2^{(\pm)}
&
\cdots
& 
\Lambda_{n}^{(\pm)}
\\
\vdots
&
\vdots
&
&
\vdots
\\
\!\!\theta_1 \Lambda_1^{n-1}\!\!
&
\!\!\Lambda_2^{(\pm)n-1}\!\!
& 
\!\!\cdots\!\!
& 
\!\!\Lambda_{n}^{(\pm)n-1}\!\!
\end{array}
\right|.
\end{eqnarray*}
By using homological relation \eqref{homological}, 
we can expand the last two factors of $Q_{n+1}$ as 
\begin{eqnarray*}
\left|
\begin{array}{ccc}
1
& 
\cdots
& 
\fbox{$1$}
\\
\Lambda_2^{(\pm)}
&
\cdots
& 
\Lambda_{n+1}^{(\pm)}
\\
\vdots
&
&
\vdots
\\
\Lambda_2^{(\pm)n-1}
& 
\cdots
& 
\Lambda_{n+1}^{(\pm)n-1}
\end{array}
\right|
\!\!\!&=&\!\!\!
\left|
\begin{array}{cccc}
1
& 
\cdots
& 
1
& 
\fbox{$0$}
\\
\Lambda_2^{(\pm)}
&
\cdots
&
\Lambda_n^{(\pm)}
& 
0
\\
\vdots
&
&
\vdots
&
0
\\
\Lambda_2^{(\pm)n-1}
& 
\cdots
& 
\Lambda_n^{(\pm)n-1}
& 
1
\end{array}
\right|
\left|
\begin{array}{ccc}
1
& 
\cdots
& 
1
\\
\Lambda_2^{(\pm)}
&
\cdots
& 
\Lambda_{n+1}^{(\pm)}
\\
\vdots
&
&
\vdots
\\
\Lambda_2^{(\pm)n-1}
& 
\cdots
& 
\fbox{$\Lambda_{n+1}^{(\pm)n-1}$}
\end{array}
\right|
\\
\left|
\begin{array}{ccccc}
\fbox{$\theta_1$}
& 
1
& 
\cdots
& 
1
\\
\theta_1\Lambda_1
&
\Lambda_2^{(\pm)}
&
\cdots
& 
\Lambda_{n}^{(\pm)}
\\
\vdots
&
\vdots
&
&
\vdots
\\
\!\!\theta_1 \Lambda_1^{n-1}\!\!
&
\!\!\Lambda_2^{(\pm)n-1}\!\!
& 
\!\!\cdots\!\!
& 
\!\!\Lambda_{n}^{(\pm)n-1}\!\!
\end{array}
\right|
\!\!\!&=&\!\!\!
\left|
\begin{array}{cccc}
\!\!\fbox{$0$}\!\!
& 
1
& 
\cdots
& 
1
\\
0
&
\Lambda_2^{(\pm)}
&
\cdots
&
\!\!\Lambda_{n}^{(\pm)}\!\!
\\
0
&
\vdots
&
&
\vdots
\\
\!\!1\!\!
&
\!\!\Lambda_2^{(\pm)n-1}\!\!
& 
\!\!\cdots\!\!
& 
\!\!\Lambda_{n}^{(\pm)n-1}\!\!
\end{array}
\right|
\left|
\begin{array}{ccccc}
\theta_1
& 
1
& 
\!\!\cdots\!\!
& 
1
\\
\!\!\theta_1\Lambda_1\!\!
&
\Lambda_2^{(\pm)}
&
\!\!\cdots\!\!
& 
\!\!\Lambda_{n}^{(\pm)}\!\!
\\
\vdots
&
\vdots
&
&
\vdots
\\
\!\!\fbox{$\theta_1 \Lambda_1^{n-1}$}\!\!
&
\!\!\Lambda_2^{(\pm)n-1}\!\!
& 
\!\!\cdots\!\!
& 
\!\!\Lambda_{n}^{(\pm)n-1}\!\!
\end{array}
\right|. 
\end{eqnarray*}
We note that 
the two quasideterminants involving $\fbox{$0$}$ element 
are exactly equal to each other by Schur complements \eqref{schur}. 
Substituting the above equations into $Q_{n+1}$, and using the fact that
$[\Lambda_i,\Lambda_j]=0$, we get 
\begin{eqnarray*}
Q_{n+1}&=&Q_n \Lambda_1-
 \Lambda_{n+1}^{(\pm)}Q_n\\
&=&
\left(
\begin{array}{cc}
(\lambda_1-\lambda_2^{(\pm)})\cdots
(\lambda_1-\lambda_{n+1}^{(\pm)})a_1e^{L_1}
&
(\mu_1-\lambda_2^{(\pm)})\cdots
(\mu_1-\lambda_{n+1}^{(\pm)})b_1e^{-\overline{L}_1}
\\
-(\lambda_1-\lambda_2^{(\mp)})\cdots
(\lambda_1-\lambda_{n+1}^{(\mp)})\overline{b}_1e^{-L_1}
&
(\mu_1-\lambda_2^{(\mp)})\cdots
(\mu_1-\lambda_{n+1}^{(\mp)})a_1e^{\overline{L}_1}
\end{array}
\right).
\end{eqnarray*}
As for $D_n$, we can use the noncommutative Jacobi identity \eqref{jacobi} and the homological relation \eqref{homological} to get
\begin{eqnarray*}
D_n=
-
\left|
\begin{array}{ccc}
1
& 
\cdots
& 
1
\\
\Lambda_2^{(\pm)}
&
\cdots
& 
\Lambda_{n}^{(\pm)}
\\
\vdots
&
&
\vdots
\\
\!\!\Lambda_2^{(\pm){n-2}}\!\!
& 
\!\!\cdots\!\!
& 
\!\!\fbox{$\Lambda_{n}^{(\pm){n-2}}$}\!\!
\end{array}
\right|
\!
\Lambda_{n}^{(\pm)}
\!
\left|
\begin{array}{ccc}
1
& 
\cdots
& 
\fbox{$1$}
\\
\Lambda_2^{(\pm)}
&
\cdots
& 
\Lambda_{n}^{(\pm)}
\\
\vdots
&
&
\vdots
\\
\!\!\Lambda_2^{(\pm){n-2}}\!\!
& 
\!\!\cdots\!\!
& 
\!\!\Lambda_{n}^{(\pm){n-2}}\!\!
\end{array}
\right|^{-1}
\!\!\!\!\!\!\!=\!
-\!
\Lambda_{n}^{(\pm)}
\left|
\begin{array}{cccc}
1
& 
\cdots
&
\!\!1\!\!
& 
\fbox{$0$}
\\
\Lambda_2^{(\pm)}
&
\cdots
& 
\Lambda_{n-1}^{(\pm)}
&
\!\!0\!\!
\\
\vdots
&
&
\vdots
&
\!\!0\!\!
\\
\!\!\Lambda_2^{(\pm){n-2}}\!\!
& 
\!\!\cdots\!\!
& 
\!\!\Lambda_{n-1}^{(\pm){n-2}}\!\!
&
\!\!1\!\!
\end{array}
\right|^{-1}
\end{eqnarray*}
By continuing the same process $n-1$ times, 
we can prove the statement \eqref{Dn}. 
 
Now let us return back to the general situation that $I$ 
is not specified to $I=1$.
So far, we have shown that the asymptotic form of $J_{n}$ 
can be expressed as \eqref{asymp_Jn} 
in terms of $Q_{n}$ of \eqref{Qn} 
and $D_{n}$ of \eqref{Dn}.  
$D_{n}$ is a constant matrix and would not affect 
the result of action density. 
Hence in the original situation for finite $L_I$, 
the asymptotic form is essentially 
$J_{n} \sim Q_{n}\Lambda_IQ_{n}^{-1}$ which   
is in a very similar form of a one-soliton solution 
with $Q_{n}=\theta_I$ (Cf. \eqref{U2_n} and \eqref{Qn}).   
Under some conditions, for instance 
$\mu_{i}=\overline{\lambda}_i$, for all $i$ (not the unique choice), 
$Q_{n}$ can be simplified in a more concise form :
\begin{eqnarray}
 Q_n=\left(
\begin{array}{cc}
\label{Qn_1 soliton}
a_I^\prime e^{L_I} & b_I^\prime e^{-\overline{L}_I} \\
-\overline{b}_I^\prime e^{-L_I} & \overline{a}_I^\prime e^{\overline{L}_I}
\end{array}
\right).
\end{eqnarray}
which is exactly in the form of one-soliton solution now. 
This is what we want to show here. 
The only difference between $Q_n$ and $\theta_I$ is  the constants $a_I^\prime, b_I^\prime$ and $a_I, b_I$
which relate to the position of the principal peak of the action density (Cf. \eqref{action_cpx_2}).  
This difference gives rise to additional position shift, 
called the phase shift.  
Now we can conclude that the action density calculated from the asymptotic form of $J_{n}$ and $Q_n$ 
is almost the same as \eqref{action_cpx_2} with $\theta=\theta_I$.  
In other word, if we consider the comoving frame with $I$-th one-soliton, the asymptotic multi-soliton inherits almost the same features from the $I$-th one-soliton except for the phase shift factor.

It is time for us to calculate the phase shift factor 
 $\Delta_{I,n}$ explicitly. 
Firstly, we take the Ultrahyperbolic space $\mathbb{U}$ for example, 
\eqref{Qn_1 soliton} is obviously satisfied because
$\mu_i=\overline{\lambda}_i$, for all $i$ (Cf. Table \ref{table}). 
Then the action density becomes
\begin{eqnarray}
\label{action_U}
 {\mbox{Tr}} F_{\mu \nu}F^{\mu \nu}=
8\left[(\alpha_I\overline{\beta}_I-\overline{\alpha}_I\beta_I)
(\lambda_I -\overline{\lambda}_I)  \right]^2
\left(2{\mbox{sech}}^2 X_I-3{\mbox{sech}}^4 X_I\right),
\end{eqnarray}
where $X_I=L_I + \overline{L}_I +\log\left|a_I^\prime/b_I^\prime\right|
=L_I + \overline{L}_I +\log\left|a_I/b_I\right|
+\Delta_{I,n}^{\scriptsize{\mathbb{U}}}$. 
The phase shift factor can be calculated by \eqref{X}.  
The result is exactly real-valued 
\begin{eqnarray*}
\Delta_{I,n}^{\scriptsize{\mathbb{U}}}=
\sum_{k=1(\neq I)}^{n}\varepsilon_k
\log 
\left|
\frac{\lambda_I-\lambda_k}{\lambda_I-\overline{\lambda}_k} 
\right|,
\end{eqnarray*}
where  $\varepsilon_k=+1$ for  case (i)
and $\varepsilon_k=-1$ for  case (ii).  

For the Euclidean signature, we can use \eqref{X} and
the reality condition $\mu_i=-1/\overline{\lambda}_i$, 
for all $i$ (Cf. Table \ref{table}) 
to calculate the phase shift, the result is also real-valued:
  
\begin{eqnarray*}
\Delta_{I,n}^{\scriptsize{\mathbb{E}}}=
\sum_{k=1(\neq I)}^{n}\varepsilon_k
\log 
\left|
\frac{\lambda_I-\lambda_k}{1+\lambda_I\overline{\lambda}_k} 
\right|.
\end{eqnarray*}
For the Minkowski signature, if we don't impose additional condition 
on $\lambda_i$ and $\mu_i$ as that in one-soliton case 
(Cf. Table \ref{table}), the phase shift factor is complex-valued in general. 
This shortcoming can be solved immediately, 
for example, we can take $\mu_i=\overline{\lambda}_i$ 
or $\mu_i=-1/\overline{\lambda}_i$ like that 
in $\mathbb{U}$ and $\mathbb{E}$, respectively. 

In summary, 
the asymptotic behavior of the solutions \eqref{U2_n} 
can be interpreted as 
$n$ intersecting soliton walls with 
phase shifts in the scattering process.
It is a well-known property for 
the KP multi-solitons, but a new insight for 
anti-self-dual Yang-Mills multi-solitons.  

\section{Unitarity of the Multi-Soliton Solutions in $\mathbb{U}$}

In this section, we discuss the unitarity of multi-soliton solutions for physical purpose. 
Recall that in Table \ref{table}, we get unitary one-soliton solutions (soliton walls) 
successfully on the Ultrahyperbolic space $\mathbb{U}$ and 
hence they could be realized as branes of three-dimensions 
in the N=2 string theory as commented in introduction. 
Therefore the multi-soliton solutions (the intersecting soliton walls) 
in unitary gauge group case
can be interpreted as $n$ intersecting branes in the N=2 string theory. 

To prove the unitarity of $n$ intersecting soliton walls, 
it suffices to verify the hermiticity of gauge fields. 
By imposing the real slice condition (Cf. Table \ref{table}) 
of Ultrahyperbolic space $\mathbb{U}$ on the gauge fields 
\eqref{gauge_f_special}, we obtain the Ultrahyperbolic version 
of gauge fields : 
\begin{eqnarray}
\label{gaugeU} 
A_{0} = - A_{2} 
=\frac{1}{2} 
(\partial_{2}J\cdot J^{-1} - \partial_{0}J\cdot J^{-1}), ~~
A_{1} = -A_{3}
=\frac{1}{2}
(\partial_{3}J\cdot J^{-1} - \partial_{1}J\cdot J^{-1}).
\end{eqnarray}
We will soon see that $J$-matrix is 
constant multiple of the special unitary matrix,  
and hence all the gauge fields $A_{\mu}$ ($\mu$= 0, 1, 2, 3) 
are anti-hermitian and traceless. 

In order to discuss it, let us define a class of $2\times 2$ matrices. 
If $P\in\mathrm{GL}(2, \mathbb{C})$ has the following form, 
\begin{eqnarray}
 P=\left(
\begin{array}{cc}
\alpha & \overline{\beta}\\
-\beta & \overline{\alpha}
\end{array}
\right),~~~\alpha,\beta\in \mathbb{C},
\end{eqnarray}
then we call $P$ the pre-$\mathrm{SU}(2)$ matrix in this paper. 
We can easily find that the pre-$\mathrm{SU}(2)$ matrix $P$ can be 
represented by the product of $\det P$ and a $\mathrm{SU}(2)$ matrix
and hence satisfies $P^\dagger P=PP^\dagger=\det P$. 
Furthermore, if two $2\times 2$ matrices $P$ and $Q$ 
are pre-$\mathrm{SU}(2)$, then $P\pm Q$, $PQ$, and 
$P^{-1},Q^{-1}$ are also pre-$\mathrm{SU}(2)$. 

By the iterative representions of quasideterminants 
(Cf.\eqref{Schur complement}), 
the $n$-soliton solution $J_n$ in \eqref{U2_n} 
is found to be a polynomial of $2\times 2$ matrices  
$\theta_i, \theta_i^{-1}, \Lambda_i, \Lambda_i^{-1}$ 
which are all pre-$\mathrm{SU}(2)$. 
Therefore $J_n$ is also pre-$\mathrm{SU}(2)$ 
satisfying $J_n^\dagger J_n=J_n J_n^\dagger=\det J_n$. 
On the other hand, we can prove that $\det J_n=\displaystyle 
\prod_{i=1}^n\vert \lambda_i\vert^2$ 
by the recursion relation:  
$\det J_n = \det(\Lambda_{n})\det J_{n-1}=
\vert\lambda_{n}\vert^2 \det J_{n-1},~\det J_0=1$ 
which comes from the explicit form \eqref{Darboux_J} 
of the Darboux transformation of $J_k~(k=1,\cdots,n)$.  

Now we can prove the gauge fields \eqref{gaugeU} 
are anti-hermitian and traceless 
because of $(\partial_\mu J_n)J_n^{-1}
=(\partial_\mu \widetilde{J}_n)\widetilde{J}_n^{-1}$
where $\widetilde{J}_n$ is a $\mathrm{SU}(2)$ matrix defined by 
$\widetilde{J}_n:= R^{-1/2} J_n$ in which $R:=\displaystyle 
\prod_{i=1}^n\vert \lambda_i\vert^2$. 
Therefore
we can conclude that the $n$ intersecting soliton walls 
can be embedded into $G=\mathrm{SU}(2)$ 
gauge theory on the Ultrahyperbolic space $\mathbb{U}$.

\section{Conclusion and Discussion}

In this paper, we discussed multi-soliton dynamics of 
anti-self-dual Yang-Mills equations by analyzing 
the action density in the asymptotic region. 
By considering a comoving frame with the $I$-th soliton, 
we proved that the entire multi-soliton distribution is asymptotically equal 
to the $I$-th one-soliton distribution except for a phase shift, 
and we also calculated the phase shifts explicitly.  
Therefore, our results can be interpreted as intersecting 
soliton walls with phase shifts in the scattering process.   
It is surprising that this behavior is quite similar to 
the case of KP soliton scattering \cite{MaSa, OhWa}, 
and it suggests the viability of Sato's formulation 
for anti-self-dual Yang-Mills equations. 
Furthermore, we proved that in the Ultrahyperbolic space $\mathbb{U}$, the $J$-matrix is unitary. Hence the multi-soliton solutions can be embedded into $\mathrm{SU(2)}$ gauge theory. This implies that there would exist 
intersecting branes of three-dimensions in the N=2 string theory. 

In our current work, we focused on the pure scattering process 
of soliton walls and excluded the case of resonance processes.
The resonance processes describe soliton wall reconnections.
Actually, the classification of
all possible soliton wall distributions 
(containing the resonance processes) 
could be put into practice by adjusting the parameters of \eqref{U2_n} 
properly (Cf. \cite{OhWa}) or by following a similar strategy
of Kodama and Williams \cite{Kodama, KoWi} 
from the viewpoint of positive Grassmannians.

On the other hand, the quasi-Wronskian solutions here might 
have a similar representations of Maya diagrams or  
tau functions in Sato's theory, and relate to 
generalized Schur functions (Cf. \cite{GKLLRT, MOS}).
Perhaps after a more comprehensive study from this perspective, 
the realization of Sato's theory 
for the anti-self-dual Yang-Mills equation version  
might be achieved even though it is still a challenging problem. 
(Cf. \cite{Sato_lecture})

The extension of integrable systems to noncommutative space-time 
is not our aim in this paper, but it is still a potentially interesting topic. 
(For reviews see, e.g. \cite{Hamanaka_AIP, Hamanaka_PS, Lechtenfeld}.)
In the previous work \cite{GHHN}, we showed 
that on the Euclidean space $\mathbb{E}$, 
noncommutative multi-soliton solutions 
of anti-self-dual Yang-Mills equations is equivalent 
to the commutative ones in the asymptotic region. 
By quite similar arguments, we can make the same conclusion 
on the Ultrahyperbolic space $\mathbb{U}$. 
This means that on $\mathbb{U}$, the behavior of three-dimensional branes 
(soliton walls) in the asymptotic region are not 
affected by the background $B$-field. 
On the other hand, the soliton equations of lower-dimensional 
integrable systems can be derived from the anti-self-dual Yang-Mills 
equations by suitable reduction procedure even when the space-time 
coordinates are noncommutative \cite{Hamanaka_NPB}. 
Therefore, the techniques presented in this paper 
could be applied to the lower-dimensional
soliton equations even in noncommutative space-time 
because quasideterminants are especially suitable 
for the description of noncommutative integrable systems.  
(e.g. \cite{EGR, GHN, GiNi07, ReRu}).
Furthermore, the asymptotic behaviors of noncommutative multi-soliton solutions 
are proved to be the same as the commutative ones 
in lower-dimensional integrable
systems (e.g. \cite{DiMH_KdV,Hamanaka_JHEP,HaOk,Paniak}), 
but the physical interpretation are still uncertain.

\subsection*{Acknowledgments}

The work of MH was supported 
by Grant-in-Aid for Scientific Research (\#16K05318).
The work of SCH is supported 
by the scholarship of Japan-Taiwan Exchange Association.




\end{document}